\documentclass[referee]{raa}
\usepackage{graphicx,times}
\usepackage{natbib}
\usepackage{amssymb,amsmath}
\usepackage[section]{placeins}
\bibpunct{(}{)}{;}{a}{}{,}

\usepackage[pagebackref=true]{hyperref}
\usepackage{booktabs}
\usepackage{orcidlink}

\begin{document}

   \title{Simulation Study on Constraining GW Propagation Speed by GW and GRB Joint Observation on Binary Neutron Star Mergers}

 \volnopage{ {\bf 20XX} Vol.\ {\bf X} No. {\bf XX}, 000--000}
   \setcounter{page}{1}

   \author{Jin-Hui Rao
   \inst{1}, Shu-Xu Yi\inst{2}\orcidlink{0000-0001-7599-0174}, Lian Tao\inst{2}, and Qing-Wen Tang\inst{1} \orcidlink{0000-0001-7471-8451}
   }

   \institute{ Department of Physics, School of Physics and Materials Science, Nanchang University, Nanchang 330031, China; {\it qwtang@ncu.edu.cn}\\
        \and
             Institute of High Energy Physics, Chinese Academy of Sciences, China; {\it sxyi@ihep.ac.cn}
}

\abstract{
Theories of modified gravity suggest that the propagation speed of gravitational wave (GW) $v_g$ may deviate from the speed of light $c$. A constraint can be placed on the difference between $c$ and $v_g$ with a simple method that uses the arrival time delay between GW and electromagnetic (EM) wave simultaneously emitted from a burst event. 
We simulated the joint observation of GW and short Gamma-Ray burst (sGRB) signals from Binary Neutron Star (BNS) merger events in different observation campaigns, involving advanced LIGO (aLIGO) in design sensitivity and Einstein Telescope (ET) joint-detected with \textit{Fermi}/GBM.
As a result, the relative precision of constraint on $v_g$ can reach $\sim 10^{-17}$ (aLIGO) and $\sim 10^{-18}$ (ET), which are one and two orders of magnitude better than that from GW170817, respectively.
We continue to obtain the bound of graviton mass $m_g \leq 7.1(3.2)\times 10^{-20}\,$eV with aLIGO (ET).
Applying the Standard-Model Extension (SME) test framework, the constraint on $v_g$ allows us to study the Lorentz violation in the nondispersive, nonbirefringent limit of the gravitational sector.
We obtain the constraints of the dimensionless isotropic coefficients $\bar{s}_{00}^{(4)}$ at mass dimension $d = 4$, 
which are $-1\times 10^{-15}< \bar{s}_{00}^{(4)}<9\times 10^{-17}$ for aLIGO and $-4\times 10^{-16}< \bar{s}_{00}^{(4)}<8\times 10^{-18}$ for ET.
}

   \authorrunning{J.-H.R et al. }            
   \titlerunning{Constraining GW speed with joint observation}  
   \maketitle

%
\section{Introduction} 
\label{sec:intro}
Gravitational waves (GW), which are perturbations created by the bending of space-time due to the distribution of mass, are one of the key predictions of Einstein's theory of General Relativity (GR). The first direct detection of a GW event, GW150914~\citep{gw150914}, was observed from a binary black hole (BBH) merger during the first observing run (O1) of Advanced Laser Interferometer Gravitational-Wave Observatory (aLIGO)~\citep{aligo}, opening a new window to astronomy and physics. The GW observations enable us to test gravity theories in strong and dynamic regimes of gravity. To test gravity theories with GWs, it is crucial to search for anomalous deviation from the prediction of GR in model-independent ways. One of the tests is measuring the propagation speed of a GW. 
The physical implications of limiting $v_g$ are manifold. 
The limit on the speed of GWs is closely related to the validity and accuracy of general relativity. 

According to GR, in the limit in which the wavelength of GW is small compared to the radius of curvature of the background spacetime, the waves propagate along null geodesics of the background spacetime~\citep{willConfrontationGeneralRelativity2014}, \textit{i.e.}, a GW propagates with the speed of light $c$ in vacuum. In the theories of modified gravity, the $v_g$ could deviate from $c$ due to the modification of gravity such as the coupling of gravitation to "background" gravitational fields. A GW generically propagates at a speed different from $c$ and with frequency dependence dispersion relations in some theories of modified gravity (see,\citep{sme-kostelecky2016testing,gravitythe-1,blas_constraining_2016,Bettoni2017,derhamGravitonMassBounds2017}). On the observation with the GW detector, the propagation final resulting in an offset in the relative arrival times of simultaneously emitted messengers (GW and neutrinos, or photons, \textit{etc.}) at the Earth~\citep{willConfrontationGeneralRelativity2014,Lombriser_2016,gw-grb170817}.

Currently, the most exact way to measure $v_g$ is, by measuring the time delay between GW and electromagnetic (EM) signals of the same astrophysical source~\citep{Cornish2017}. 
With the Advanced Virgo~\citep{avirgo} detector joining in later O2~\citep{LVCGWTC1}, the coalescence of a binary neutron star (BNS), GW170817~\citep{LVC:gw170817, gw-grb170817} were observed by the three-detector network~\citep{LVK-network}, which is a significant observational discovery in the field of multi-messenger astronomy. In \cite{gw-grb170817}, LIGO and Virgo Scientific Collaboration (LVC) has constrained the difference between $v_g$ and $c$ to be between ($-3\times 10^{-15}c,\,+7\times 10^{-16}c$) by using the distance to the astrophysical source (the identification of NGC 4993 as the host galaxy of GW170817 and GRB170817A~\citep{coulter2017swope, coulter2017ligo, abbott2017bnsm}) and the observed time delay of ($+1.74\pm 0.05$)\,s between GRB170817A and GW170817. 
The observation of GW170817 also takes significant scientific discoveries that have set stringent limits on fundamental physics~\citep{bakerStrongConstraintsCosmological2017,lvc-TestsGeneralRelativity2019} and the origin of short Gamma-Ray Bursts (more described in detail about the observations of GW170817 and GRB 170817A, see~\citep{LVC:gw170817,detailGRB170817A}), demonstrating the importance of multi-messenger astronomy.

The LIGO-Virgo-KAGRA (LVK) 's O4 has been officially launched and has been running for the last few months. More scientific merits in fundamental physics, astrophysics, and cosmology can be excavated from a population of joint GW/EM counterparts detection. 
It also facilitates the more stringent constraints on $v_g$. 
According to scientific observation runs LVC O1-O3~\citep{LVCGWTC1,LVCGWTC2.1,LVCGWTC3} and simulative studies (such as~\citep{gwtoolboxIII}), such a joint detection of GW/GRB are rare (one out of $\sim10$ BNS merger events).
 In this paper, we present a simulation study on constraining $v_g$, using \texttt{GWToolbox}\footnote{Gravitational Wave Universe Toolbox: a Python package to simulate gravitational wave detections with different backgrounds. The web application: \url{https://gw-universe.org/}, Python module: \url{https://bitbucket.org/radboudradiolab/gwtoolbox/src/master/}}~\citep{gwtoolbox, gwtoolboxIII} to simulate the joint GW-GRB detection of a BNS merger event with both second and third-generation ground-based GW detectors, including aLIGO in design sensitivity~\citep{aligo} (design aLIGO) and Einstein Telescope (ET)~\citep{EinsteinTelescope_2010}.
 In comparison to similar simulative work in \cite{nishizawa2016}, we simulated the parameterized BNS population, which is consistent with the GWTC-3~\citep{LVCGWTC3} catalogue.

We continue to place constraints on Lorentz invariance violation.
A theoretical framework known as the Standard Model extension (SME) provides a comprehensive test framework for constraints on local Lorentz violation. The coefficients for Lorentz violation in the gravitational sector of the SME cause the modification of the group velocity of GWs. Some early work can see~\citep{sme-kostelecky2004,lv-5,sme-colladay1998,sme-kostelecky2006,sme-kostelecky2016testing}. The constraint of $v_g$ allows constraining the nine coefficients for Lorentz violation in the limit of non-birefringent, non-dispersive at mass dimension $d=4$. Thus, we use this bound to place a constraint on local Lorentz invariance violation within the context of the effective-field-theory-based test framework provided by the SME. 
Another way in which the speed of GWs could differ from $c$ is if gravitation were propagated by a massive field (a massive graviton). The massive particle has a dispersion relation, $E^2=m_g^2+p^2$ (in natural unit). Therefore, the bound on $v_g$ can be converted to a bound on the mass of graviton $m_g$~\citep{nishizawa2014}.

The structure of this paper is as follows. In Sec.~\ref{sec:method}, we discuss the methods used to calculate the bound of $v_g$, and Sec.~\ref{sec:result} presents the constrain results of GW speed. In Sec.~\ref{sec:liv} and Sec.~\ref{sec:graviton} we utilize the result of the GW speed to constrain both the coefficients for Lorentz violation and the graviton mass.

\section{Method} 
\label{sec:method} 

In this section, we will discuss the methodologies employed to obtain constraints on $v_g$, as well as our approach to simulating the observation on a BNS merger population using the \texttt{GWToolbox}.

\subsection{Propagation Speed of GW}
\label{speed of gravity}
In the following, we are going to present a general method to constrain $v_g$. We will start with a brief introduction to the method, comparing the arrival times between GW and a high-energy photon from a sGRB~\citep{gw-grb170817} to measure the $v_g$. 

Considering the compact binaries (like BNS) within the cosmological distances, assuming there is a small discrepancy $\delta t$ between the propagation time of GWs and EM waves emitted by the sources, after traversing a distance of $D$. The fractional speed difference between GWs and EM waves (or photons) during the trip can be simply expressed as~\citep{gw-grb170817}
\begin{equation}
    \frac{\delta c}{c} \approx c \frac{\delta t}{D},
\label{eq: easy one}
\end{equation}
we define $\delta c \equiv v_{g}-c$, is the difference between $v_{g}$ and $c$, the time decay $\delta t  \equiv \Delta t_{\rm{obs}}-\Delta t_{\rm{int}}$, where $\Delta t_{\rm{obs}}$ and $\Delta t_{\rm{int}}$ are the differences we observed in the arrival time and the intrinsic emission time difference of the two signals, respectively. Assume that the refractive index of the vacuum is unity, and both waves are expected to be affected by background gravitational potentials in the same way. This relation exhibits better constrain when considering sources at larger distances. However, for sources at cosmological distances ($z>0.1$), the impact of cosmological effects cannot be disregarded. Third-generation ground-based GW detector such as ET enables us to observe NS-NS binaries at cosmological distances up to $z\sim 2$, and up to $z\sim 4$ for NS-BH binaries~\citep{EinsteinTelescope_2010}.

Now we extend the scenario to sources at cosmological distances, taking into account the effects of cosmological redshift and the influence of a specific cosmological model.
Here we refer to the approach in \cite{nishizawa2014} and \cite{nishizawa2016} that considers the case of propagation over cosmological distances. Consider a background in a flat lambda cold dark matter ($\Lambda$CDM) universe for simplicity. The comoving distance from the Earth to a source at redshift $z$ can be written as
\begin{equation}
    \chi (0,z)=\int_{z_0=0}^{z} \frac{v_g}{H(z)}\mathrm{d}z \equiv (1-\frac{\delta c}{c})\chi _0(0,z),
\end{equation}
the Hubble parameter $H(z)$ is given by
\begin{equation}
    H(z)=H_0\sqrt{\Omega_m(1+z)^3+\Omega_{\Lambda}},
\end{equation}
where $\Omega_m$ and $\Omega_{\Lambda}$ are the matter density parameter and the cosmological constant, respectively, and $H_0$ is the Hubble constant. Here use the cosmological parameters of the cosmos from Planck18~\citep{Planck:2018vyg}, which are $H_0=67.66\, \mathrm{km\,Mpc^{-1}\,s^{-1}}
$, $\Omega_m=0.30966$. We express $\chi_0(0,z)$ as the comoving distance when $v_g=c$.
Consider a scenario in which GW and EM signals are emitted from the same source of redshift $z$ with velocities of $c$ and $v_g$, respectively. 
Both signals travel the same comoving distance from the source, yet they reach us at redshifts $z=0$ and $z=-\Delta z$, respectively.
Neglecting the second-order (and more) correction in $\Delta z$ (more detailed derivations are given in \cite{nishizawa2014} and \cite{nishizawa2016}), for $\chi(0,z+\Delta z)=\chi_0(0,z)$ gives the time delay $\delta t$ between GWs and photons introduced by $\delta c/c$ is~\citep{nishizawa2014}
\begin{equation}
    \delta t=\frac{\Delta z}{H_0}=\frac{\delta c}{c}\int_{0}^{z}\frac{\mathrm{d} z}{H(z)}.
\label{eq: cosmosdistance}
\end{equation}
In addition, after redshifting the intrinsic time delay, the time delay in propagation $\delta t$ becomes:
\begin{equation}
    \delta t  = \Delta t_{\rm{obs}}-(1+z)\Delta t_{\rm{int}},
\label{eq: timerelation}
\end{equation}
where $z$ is the redshift of the source.

Here we use Equation~\eqref{eq: cosmosdistance} to obtain the fractional speed difference $\delta c/c$ for any sources that know the distance or redshift and the time delay between the GW signal and EM signal in propagation. While the $v_g$ is time-dependent in general~\citep{speed-1,speed-3}, and may deviate in varying scales from the speed of photons through coupling with the different gravitational fields during propagation in the current epoch of the Universe (if there is any modification of gravity which allows for a change in the GW propagation speed), in our work, we regard the speed of the GW as constant during propagation for simplicity, or rather, the $v_g$ being measured here represents the average speed during a trip.

\subsection{Simulating the Observation on a BNS Merger Population}
As in the observations~\citep{gw-grb170817} of GW170817 and GRB70817A, in our simulations, it is assumed that the GW signal arrives at Earth before the GRB signal.
To obtain more data on joint events that were heard (GW signal) and seen (EM signal) later, and to strengthen further constraints on the bounds of $\delta c/c$, 
we use \texttt{GWToolbox} to simulate the joint detection of GW and GRB by design aLIGO and ET on a BNS merger population, using the same cosmological model used in Equation~\eqref{eq: cosmosdistance}. The GRB detector for simulating joint detection here utilizes only the \textit{Fermi}/GBM. $\Delta t_{\rm{obs}}$ is explained to be the time difference between the peak (usually in the merger phase) of the GW signal and the first sGRB photon in the propagation. 
Using the joint observation module of the \texttt{GWToolbox}, after setting a campaign of detection, one can get the catalogue that includes a list of parameters of the samples where the BNS merger is successfully joint-detected by both the specified GW and GRB detectors in all simulated GW event samples. This includes the masses of the binary, cosmological redshift, luminosity distance, single spin parameter, \textit{etc.}, and the uncertainties of corresponding parameters. 
The detection rates of our simulation correspond to a certain GRB population model. The population model and its hyper-parameters are followed from \cite{gwtoolbox} and \cite{gwtoolboxIII}, which can successfully reproduce the detection rate consistent with historical GW and GRB observations (GWTC-3 and \textit{Fermi}/\textit{Swift} GRB catalogs).

There is no capability currently for simulation of the arrival time delay between GW and sGRB signal by \texttt{GWToolbox}. Also, in our simulation, the sGRBs signal detected in joint detection with the GWs signal is on-axis. 
Although, observationally, the $\Delta t_{\rm{obs}}$ of GW170817/GRB170817A is $1.7\,$s, some studies (like~\citep{gw-grb170817}) suggest that its corresponding sGRB is off-axis, and $\Delta t_{\rm{obs}}$ is supposed to be smaller for on-axis cases.
Therefore, in this work, we assume that the statistical fluctuation of the arrival time delay $\Delta t_{\rm{obs}}$ is approximated by a truncated normal distribution between $0.5\,$s and $1.5\,$s with a mean $1\,$s, and a standard deviation of $0.5\,$s.

\textit{Realization}------After establishing a set of campaign parameters such as the signal-to-noise ratio (SNR) threshold of GW detection, observation time, and GW detector, a catalogue of samples is simulated. 
A set of $\Delta t_{\rm{obs}}$ is then randomly assigned, which is conveniently regarded as a "Realization" (\textit{i.e.}, simulating a realistic outcome once).
Table~\ref{tab:one realization} displays an example of a realization based on BNS merger events joint-detected by design aLIGO and \textit{Fermi}/GBM over a 10-year observation period with an SNR threshold of 8.0. Here we only show the important parameters used for calculation in the table. The complete catalogue also includes a single effective spin parameter, the uncertainty of masses of the binary stars, and so on.
We can see that for each detected BNS merger event in Table~\ref{tab:one realization}, there is a redshift $z$ and the corresponding uncertainty $\mathrm{d}z$, as well as the arrival time delay $\Delta t_{\rm{obs}}$. 
The uncertainties $\mathrm{d}z$ listed in Table~\ref{tab:one realization} are estimated from those of the luminosity distance $\mathrm{d}L$ of the GW observation, which are calculated with a Fisher information matrix method with the \texttt{GWToolbox}~\citep{gwtoolbox}.
By applying the Monte Carlo (MC) algorithm to sample in the parameter space of $z$ and $\Delta t_{\rm{obs}}$ for each event, we obtain the distribution of the upper (lower) bounds.
Both parameters are subject to normal distribution, 
with mean values taken from the catalogue and standard deviations based on $\mathrm{d}z$ and 0.01\,s (which is considered as the uncertainty of the merger instant inferred from GW observation), respectively.
Combining all events in the catalogue, we obtain the final bounds on $\delta c/c$ using Equation~\eqref{eq: cosmosdistance} and Equation~\eqref{eq: timerelation} with this realization.
Then we perform a new realization to obtain the new sample results of the catalogue in the same observation setting to constrain the $\delta c/c$.
Instead of Bayesian inference used by \cite{nishizawa2016}, here we are using a simpler MC simulation, which allows us to effectively explore many realizations and different observational campaigns.
\begin{table}[h]
\caption{One realization based on BNS merger events joint-detected by design aLIGO and \textit{Fermi}/GBM over a 10-year observation period with SNR threshold = 8.0.}
\label{tab:one realization}
\renewcommand{\arraystretch}{1.5} 
\centering
\begin{tabular}{*{7}{c}}
\toprule
 & $z$ & $dz$ & $m1$ & $m2$ & $\Delta t_{\text{obs}}$ & ... \\
\midrule
1 & 0.11 & 0.05 & 1.76 & 1.79 & 1.04 & ... \\
2 & 0.06 & 0.02 & 1.96 & 1.22 & 0.99 & ... \\
3 & 0.08 & 0.04 & 1.60 & 1.61 & 1.00 & ... \\
4 & 0.06 & 0.03 & 1.31 & 2.24 & 0.85 & ... \\
5 & 0.07 & 0.03 & 1.42 & 2.47 & 1.25 & ... \\
\bottomrule
\end{tabular}
\end{table}
 
Taking into account the contingency of the distance of the detected events in a simulation,
it is essential to conduct multiple realizations of a simulated observation setting. 
We call it a successful realization if there is at least one BNS merge event in which both the GW and the GRB are detected. After 100 realizations are successfully performed, the results of all realizations are combined to obtain the final constraint interval of $\delta c/c$.

\begin{figure*}[ht]
    \centering
    \includegraphics[width=1\linewidth]{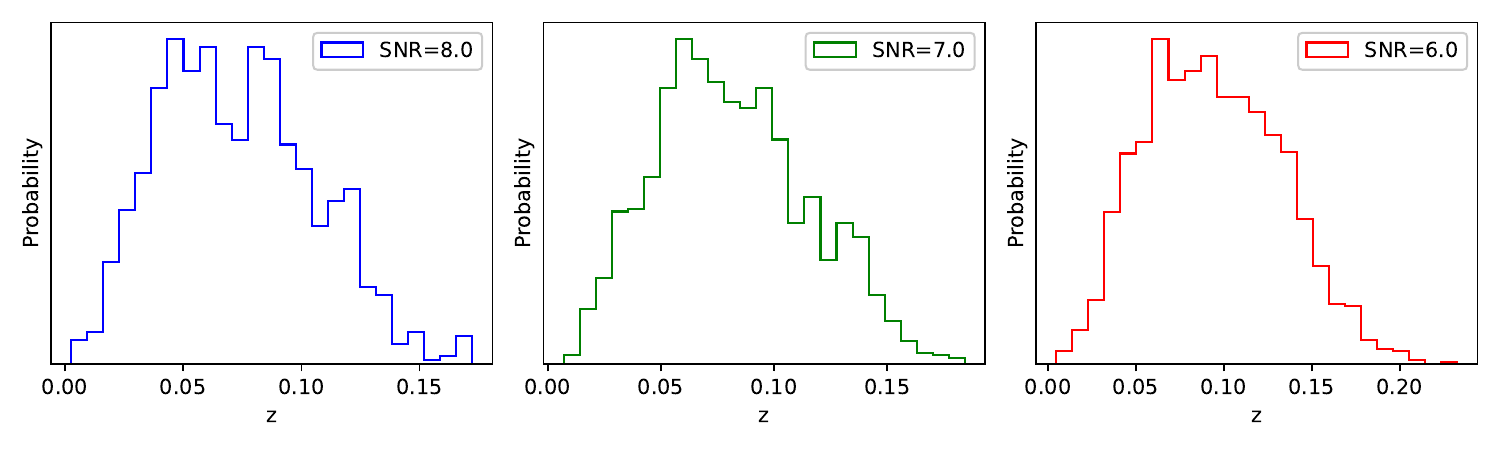}
    \includegraphics[width=1\linewidth]{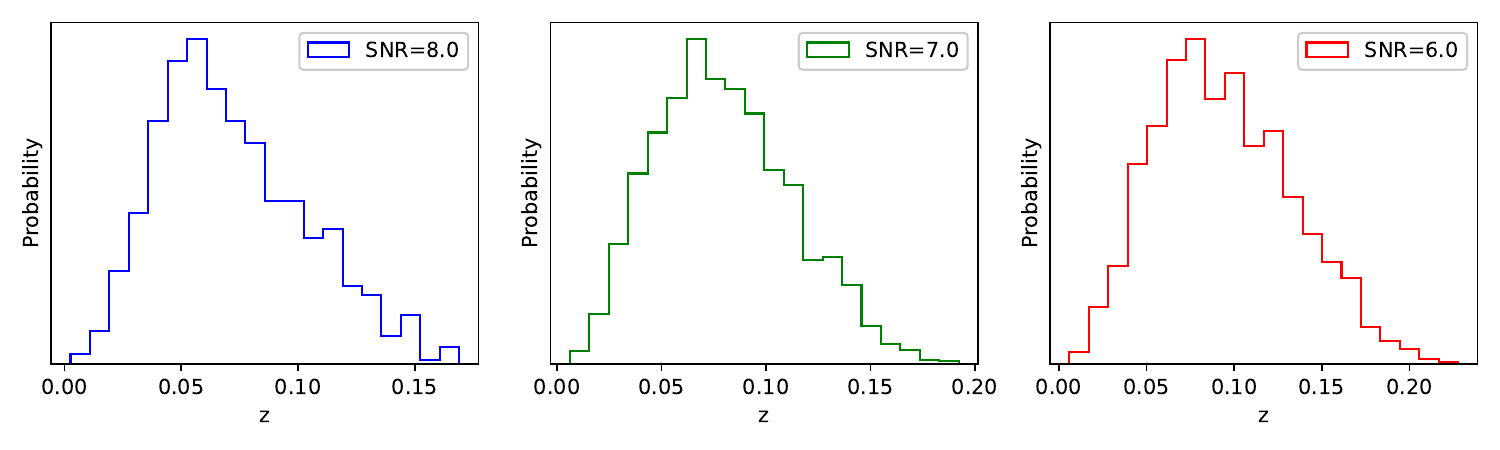}
    \caption{The $z$ distributions of simulated catalogue that are averaged over all realization, is presented in the top three panels for design aLIGO over a 10-year observation period with SNR thresholds of 8.0, 7.0, and 6.0. The bottom three panels display the distribution for design aLIGO over a 20-year observation period.}
    \label{fig:z_dist_ligo-20y}
\end{figure*}

\begin{figure*}[ht]
    \centering
    \includegraphics[width=1\linewidth]{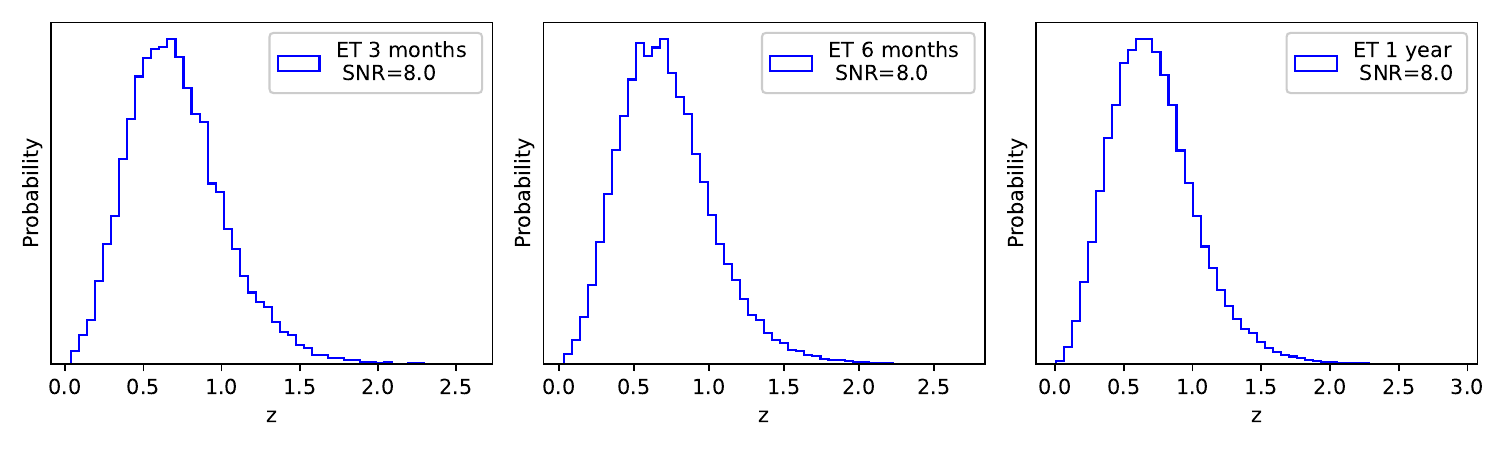}
    \caption{The $z$ distributions of simulated catalogue that are averaged over all realization, which is observed by ET over different observation times with SNR threshold of 8.0.}
    \label{fig:z_dist_et-snr8-all}
\end{figure*}

The redshift distribution of samples that combines all realizations is shown in Figure~\ref{fig:z_dist_ligo-20y}, which uses design aLIGO over an observation period of 10/20 years with different values of SNR threshold. Figure~\ref{fig:z_dist_et-snr8-all} illustrates the redshift distribution that combines all realizations from ET over three observation times of 3 months, 6 months, and 1 year with SNR threshold = 8.0. As depicted in the distribution, nearly 20$\%$ of the samples in design aLIGO observations have a redshift above 0.1, and most of the samples are greater than 0.5 for ET. Therefore, the measurement of the GW speed of all events takes into account cosmological effects.

\section{Result on $\delta c/c$}
\label{sec:result}

\textit{the upper bound and the lower bound of $\delta c/c$}------ the true intrinsic time difference range between GW and sGRB is unknown.
To obtain the upper bound of $\delta c/c$, we conservatively assume that the peak of the GW signal and the sGRB were emitted simultaneously, \textit{i.e.}, to consider the intrinsic time delay is 0, attributing the entire observed time delay to faster propagation of GW. 
For the lower bound, a conservative bound relative to the few seconds of intrinsic time delays has been discussed in~\citep{gw-grb170817,nishizawa2014,timelag_Li_2016}. We conservatively consider the intrinsic time delay to be 10 seconds, \textit{i.e.}, the sGRB signal was emitted 10 seconds after the GW signal. 
\footnote{The dispersion of the intergalactic medium (like plasma) has a negligible effect on the gamma-ray photon, with an expected propagation delay many orders of magnitude smaller than our assumption for $\Delta t_{int}$~\citep{zhangFRB,gw-grb170817}.}

As discussed above, after performing one realization, we obtain a conservatively best constraint interval that combines all the events in the catalogue. 
Combining all realizations then gives the final constraint range. 
To place a constraint on $\delta c/c$ by combining all events joint-detected by design aLIGO and \textit{Fermi}/GBM over an observation time of 20 years, the best bound (when SNR threshold is 6.0) is (90\% confidence interval)
\begin{equation}
-1.37^{+0.15} _{-0.20} \times 10^{-16}\leq \frac{\delta c}{c} \leq +9.40^{+3.40} _{-2.45} \times 10^{-18}.
\label{eq: result of bound}
\end{equation}
For SNR threshold=8.0, the number of events detected per realization decreases.
We obtained the constraints $(-1.73^{+0.27} _{-0.31}\times 10^{-16}, 1.31^{+0.57} _{-0.35}\times 10^{-17})$ of $\delta c/c$. Meanwhile, for ET, the bound over a time of 1 year (SNR threshold=8.0):
\begin{equation}
-5.11^{+0.01} _{-0.02} \times 10^{-17}\leq \frac{\delta c}{c} \leq +1.08^{+0.14} _{-0.15}\times 10^{-18}.
\label{eq: result of bound2}
\end{equation}

The results above indicate that the constraint on $\delta c/c$ has the potential to be about 10 times better than that of GW170817 after 20 years of observation by design aLIGO.
Additionally, the constraint from ET in one year of observation time is only one order of magnitude stronger than the result calculated by design aLIGO in 20 years, and two orders of magnitude stronger than GW170817.
All of the constraints (90\% confidence level) can be seen in Table~\ref{tab:ligo-res} and Table~\ref{tab:et-res}.
\begin{table}[htb]
\caption{The constraints of design aLIGO joint-detected with \textit{Fermi}/GBM in 10/20 years observation time with different SNR thresholds.}
\label{tab:ligo-res}
\renewcommand{\arraystretch}{1.5} 
\centering
\begin{tabular}{ccc}
\toprule
 Observation time & SNR threshold & Constraint of $\delta c/c$ \\
\midrule
10 years & 8.0 & $(-1.88\times 10^{-16}\,, +1.51\times 10^{-17})$ \\
10 years & 6.0 & $(-1.49\times 10^{-16}\,, +9.99\times 10^{-18})$ \\
20 years & 8.0 & $(-1.73\times 10^{-16}\,, +1.31\times 10^{-17})$ \\
20 years & 6.0 & $(-1.37 \times 10^{-16} \,, +9.40 \times 10^{-18})$ \\ \bottomrule
\end{tabular}
\end{table}
\begin{table}[h]
\caption{The constraints of ET joint-detected with \textit{Fermi}/GBM in different observation time with an SNR threshold of 8.0.}
\label{tab:et-res}
\renewcommand{\arraystretch}{1.5} 
\centering
\begin{tabular}{ccc}
\toprule
Observation time & SNR threshold & Constraint of $\delta c/c$ \\
\midrule
3 months & 8.0 & $(-5.13\times 10^{-17}\,, +1.27\times 10^{-18})$ \\
6 months & 8.0 & $(-5.11\times 10^{-17}\,, +1.15\times 10^{-18})$ \\
1 year & 8.0 & $(-5.11\times 10^{-16}\,, +1.08\times 10^{-18})$ \\
\bottomrule
\end{tabular}
\end{table}

\section{Local Lorentz Violation} 
\label{sec:liv}

SME is a comprehensive and effective ﬁeld theory that characterizes general Lorentz violations, which contains both general relativity and the Standard Model of particle physics. As a realistic theory, it can be applied to analyze observational and experimental data. The early foundational work can be found in~\citep{sme-colladay1998}, while an updated review of observational and experimental results can be seen in~\citep{sme-kosteleckyDataTablesLorentz2011}. Additionally, some early study~\citep{sme-kostelecky2004, sme-kostelecky2016testing, sme-kostelecky2006} showcases the foundational work in the gravitational sector of the SME.
The SME provides a general test framework for testing Lorentz invariance.
One advantage of interpreting constraints on Lorentz invariance violation in terms of limits on the coefficients of the SME over model-independent tests is that it allows for direct comparison of results from different kinds of experiments (e.g. polarization and time-of-flight measurements)~\citep{sme-Kislat2015}.
A Lorentz-violating term in the Lagrange density of the SME is an observer scalar density formed by contracting a Lorentz-violating operator with a coefﬁcient that acts to govern the term. The operator can be characterized in part by its mass dimension $d$, which determines the dimensionality of the coefﬁcient~\citep{sme-Kostelecky2009}.

In a vacuum, Lorentz symmetry implies that all massless waves propagate at the speed of light. However, Lorentz symmetry is spontaneously violated when a medium is present, and propagation speeds can differ. 
Within a comprehensive effective ﬁeld theory description of Lorentz violation~\citep{sme-colladay1998,sme-kostelecky2015,sme-Colladay1997}, there will be a Lorentz violation in both the gravitational and photon sectors of the SME. However, the two sectors of the violation are different, reflected in the effective field theory operator by the fact that the coefficients of the Lorentz violation factor are different. 
The difference in the group velocity of GWs and EM waves is controlled by differences in coefﬁcients for Lorentz violation in the gravitational sector and the photon sector at each mass dimension $d$.
The coefficients for Lorentz violation can be described by an observer scalar and hence can be expanded in terms of spherical harmonics~\citep{sme-kostelecky2015,sme-kostelecky2016testing}. 
At each ﬁxed $d$, there are $(d-1)^2$ independent spherical coefficients, which are used to express the Lorentz-violating degrees of freedom.

In the nondispersive, nonbirefringent limit at mass dimension $d=4$ (since every coefficient for Lorentz violation with $d > 4$ is associated with powers of momenta in the dispersion relation, only coefficients with $d = 4$ produce dispersion-free propagation~\citep{sme-kostelecky2004}), 
using natural units and assuming that the nongravitational sectors, including the photon sector, are Lorentz invariant, attributing the difference in the group velocity to the Lorentz violation from the gravitational sector. 
In this case, the modified group velocity of $v_g$ takes the form~\citep{sme-kostelecky2015,sme-Colladay1997,sme-colladay1998,sme-kostelecky2016testing}:
\begin{equation}
    v_g=1+\frac{1}{2}\sum_{\ell m}(-1)^{\ell} Y_{\ell m}(\alpha,\delta)\bar {s}_{\ell m}^{(4)},
\label{eq: liv}
\end{equation}
where $Y_{\ell m}$ is a basis of spherical harmonics, in which $\ell \leq d-2=2$, the spherical-basis coefficients $\bar {s}_{\ell m}^{(4)}$ reflects the nine Lorentz-violating degrees of freedom for Lorentz violation in the gravitational sector.
The direction $(\alpha,\delta)$ refers to the sky position of the source.
For the difference in group velocities, we have 
\begin{equation}
    \Delta v=- \sum_{\ell m,}Y_{\ell m}(\alpha,\delta)\Big ( \frac{1}{2}(-1)^{1+\ell}\bar{s}_{\ell m}^{(4)}  \Big ).
\label{eq: liv-last}
\end{equation}

To place the constraints, we have proceeded under a simplified approach called a maximum-reach approach~\citep{mra-Natasha2017} as used in \cite{gw-grb170817}, separately considering each component in $\bar{s}_{\ell m}^{(4)}$ and setting all other coefﬁcients to be zero at one time.
In the context of the maximum reach approach, sometimes referred to as a coefficient separation approach, a series of nine coefficients in Equation~\eqref{eq: liv-last} are constrained one at a time using a single measurement in \cite{gw-grb170817}.
For the purpose of this demonstration of the principle, only the first isotropic item $\bar{s}_{00}^{(4)}$ is constrained in this paper, where the term is not associated with the sky direction of the event source. Calculating this leading term is sufficient for our purposes.
The constraint of $\bar{s}_{00}^{(4)}$ with different GW detectors (joint-detected with \textit{Fermi}/GBM) are as follows:
\begin{equation}
\begin{aligned}
   \mathrm{design\ aLIGO} :& -1\times 10^{-15}< \bar{s}_{00}^{(4)}<9\times 10^{-17},\\
   \mathrm{ET}:& -4\times 10^{-16}< \bar{s}_{00}^{(4)}<8\times 10^{-18}.
\end{aligned}
\end{equation}
These constraints are obtained from the observation period of 20 years (SNR threshold=8.0) for design aLIGO and 1 year (SNR threshold=8.0) for ET.
For design aLIGO, the isotropic upper bound of $\bar{s}_{00}^{(4)}$ shows a 1 order of magnitude improvement compared to the constraint from GW170817~\citep{gw-grb170817}, while for ET, it shows a 3 orders of magnitude improvement in only 1 year observation period.

\section{Bound on Graviton Mass} 
\label{sec:graviton}

In theories of modified gravity (as described in~\citep{derhamGravitonMassBounds2017}), GWs generically propagate at a speed that differs from c and with frequency dependence dispersion relations. 
\cite{graviton3-PhysRevD.57.2061} describes that in the case of inspiralling compact binaries, GWs emitted at low frequency early in the inspiral will travel slightly slower than those emitted at high frequency later, resulting in an offset in the relative arrival times at a detector. This modification affects the phase evolution of the observed inspiral gravitational waveform, similar to that caused by post-Newtonian corrections to quadrupole phasing. Matched filtering of the waveforms could limit such frequency-dependent variations in propagation speed and thereby constrain the graviton mass.

If gravitation were propagated by a massive field, meaning the graviton is a massive particle, in which case the group velocity would be given by $v\equiv \partial E/\partial p$, in a local inertial frame, 
\begin{equation}
    \frac{v_g^2}{c^2}= 1-\frac{m_g^2c^4}{E^2}.
\end{equation}
For the relativistic regime $m\ll E$, or in other words, if the frequency of the GWs satisfied $h f\gg m_gc^2$, where $h$ is Planck’s constant, then the velocity of GWs (gravitons) will depend upon their frequency as shown in~\citep{willConfrontationGeneralRelativity2014}:
\begin{equation}
    \frac{v_g}{c} \approx 1-\frac{1}{2}(\frac{c}{\lambda _gf})^2,
\label{eq: graviton}
\end{equation}
where $\lambda _g=h/m_gc$ is the Compton wavelength of graviton. Therefore the bounds on speed of GW $v_g$ measured above are converted to the bound on graviton mass. 

\begin{figure}[ht]
    \centering
    \includegraphics[width=1\linewidth]{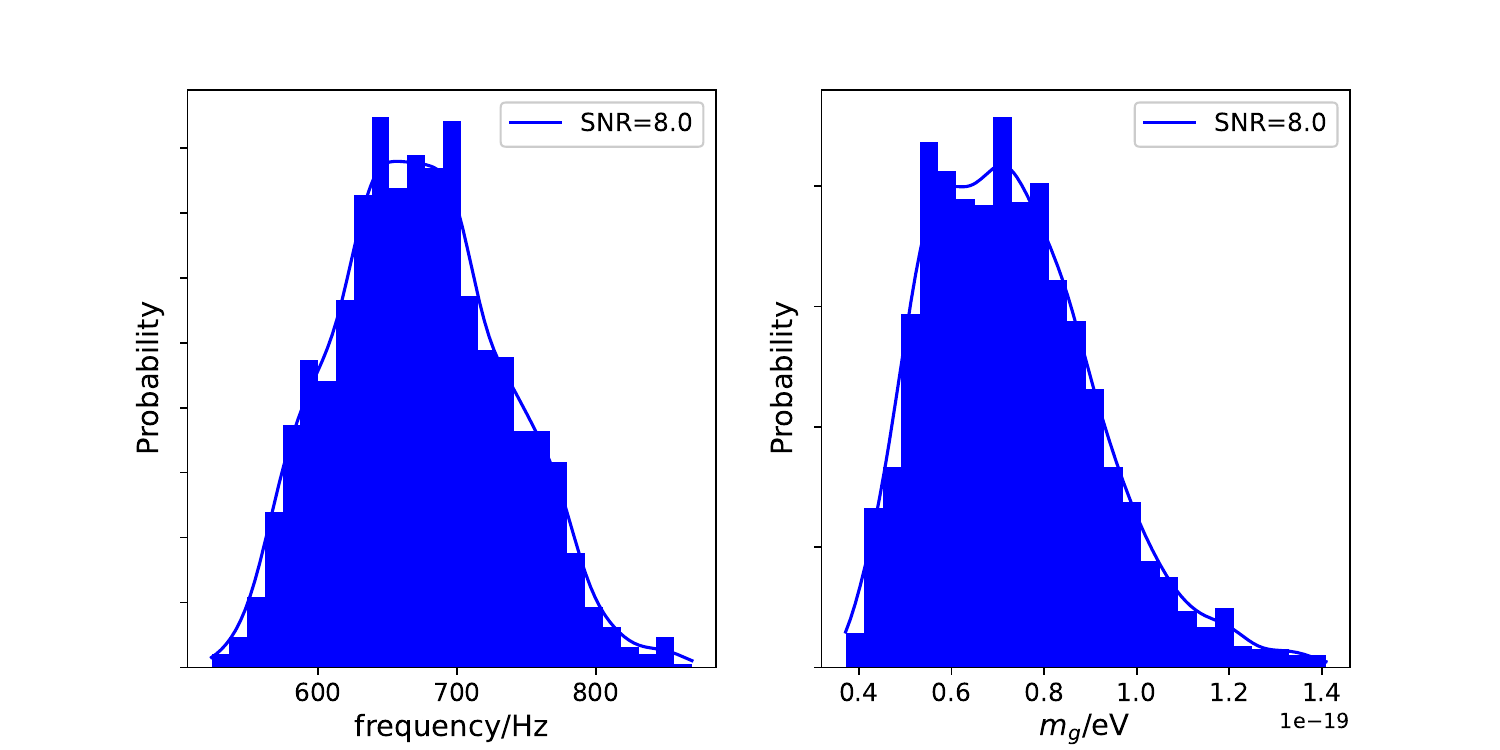}
    \includegraphics[width=1\linewidth]{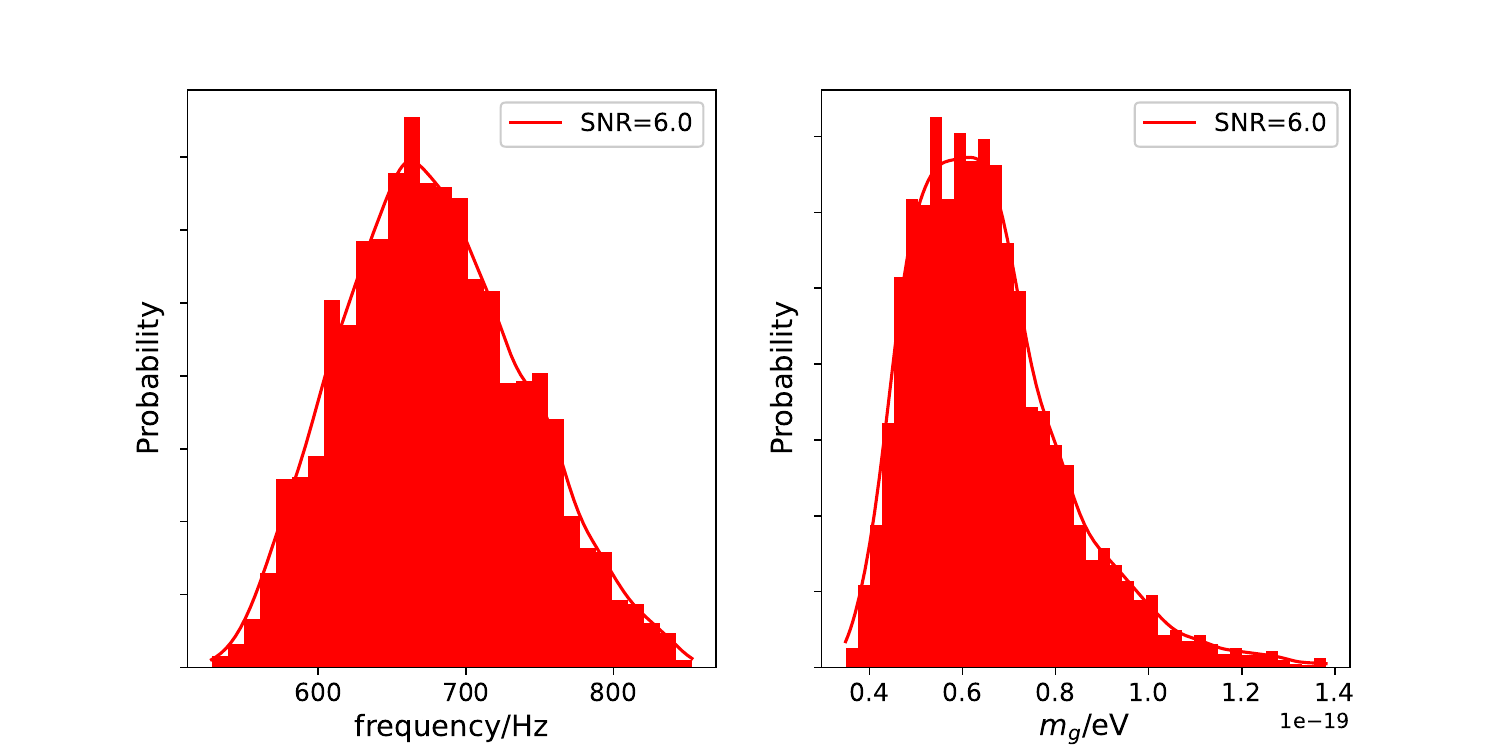}
    \caption{The frequency distribution of samples and the corresponding computed graviton mass distribution, from design aLIGO with SNR threshold equals 8.0 (top two panels) and 6.0 (bottom two panels) in the observation period of 20 years. }
    \label{fig:freq_m_g_ligo_20y}
\end{figure}

\begin{figure}[ht]
    \centering
    \includegraphics[width=1\linewidth]{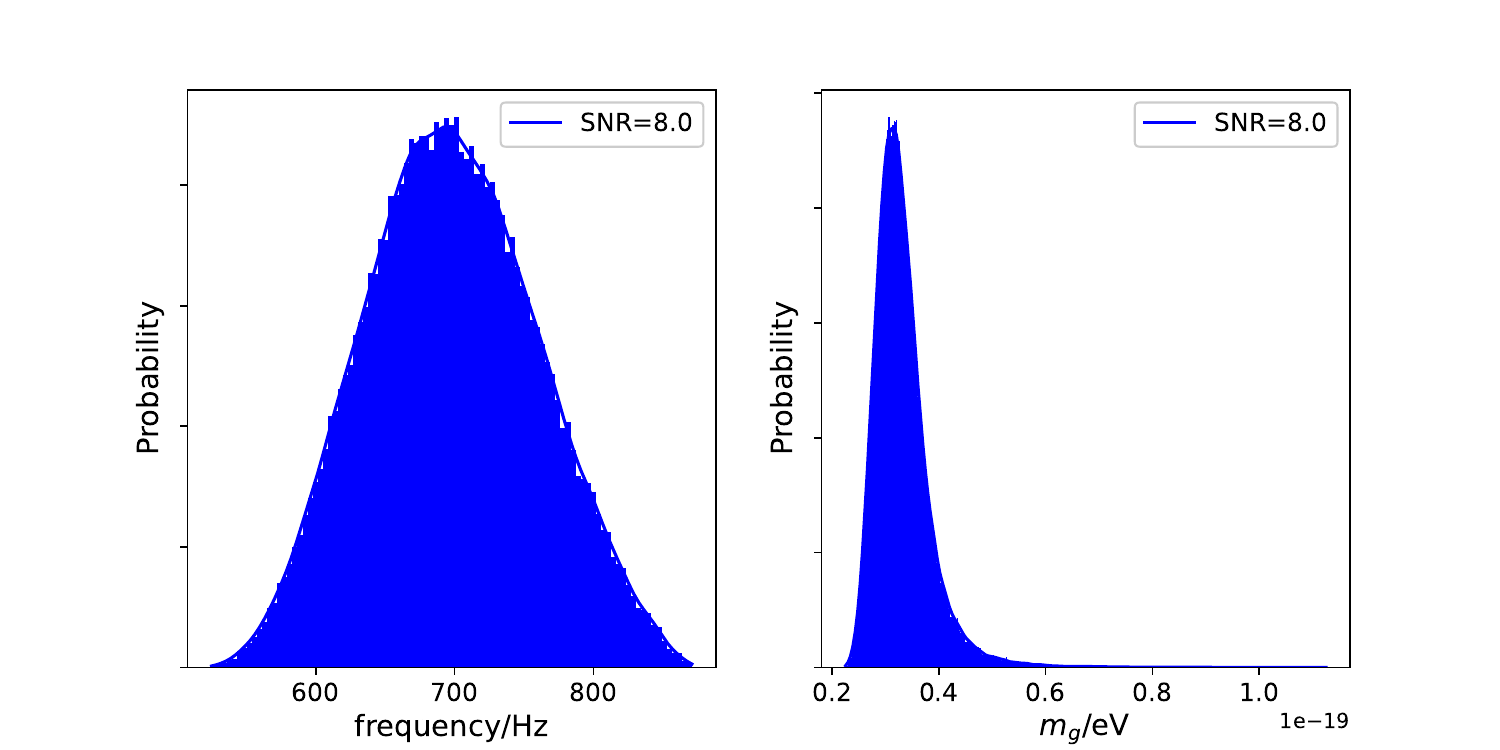}
    \caption{The frequency distribution of samples and the corresponding computed graviton mass $m_g$ distribution, from ET with SNR threshold equals 8.0 in the observation period of 1 year.}
    \label{fig:freq_m_g_et_1y}
\end{figure}

In the frequency range centered around 100\,Hz, the strain sensitivity of the design is ten times better than of the initial LIGO~\citep{aligo}. The signal of GW170817 detected by LIGO-Virgo Collaboration (LVC) has a frequency between approximately 20 and 350\,Hz~\citep{LVC:gw170817},
and the ET is designed to be 10 times more sensitive than advanced ground-based detectors, covering a frequency range of 1 to $10^4\,$Hz~\citep{EinsteinTelescope_2010}.
The limit on the GW speed corresponds to a narrow frequency range of GW in the merger stage.
The corresponding frequency can be described by the following equation~\citep{GWVol1-Maggiore:2007ulw}:
\begin{equation}
    f_{\rm{gw}}(\tau ) \simeq 134\mathrm{Hz}(\frac{1.21M_{\odot }}{M_{c}} )^{5/8}(\frac{1\mathrm{s} }{\tau } )^{3/8},
\label{eq: freq}
\end{equation}
where the $\tau$ is the time to coalescence, and $M_c$ is the chirp mass of binaries. Equation~\eqref{eq: freq} gives the GW frequency corresponding to $\tau$ seconds before coalescence, with $\tau = 0.01\,$s as the uncertainty of the merger instant inferred from GW observation, or more specifically, the uncertainty of the peaking time of the GW signal.
For each simulated GW source in the catalogue, the corresponding binary mass data will be given. Substituting this data into Equation~\eqref{eq: freq}, we calculate the corresponding frequency for all samples from different observation campaigns. By then substituting the calculated frequency into Equation~\eqref{eq: graviton}, we obtain the distribution of $m_g$.
Figure~\ref{fig:freq_m_g_ligo_20y} and Figure~\ref{fig:freq_m_g_et_1y} show the frequency distribution and the corresponding computed graviton mass $m_g$ distribution of samples from design aLIGO and ET, respectively, which combine all realizations.

For design aLIGO in the observation period of 20 years, we obtain the constraint $m_g \leq 6.3 \times 10^{-20}\, \mathrm{eV}$ (SNR threshold=6.0) and $m_g \leq 7.1 \times 10^{-20}\, \mathrm{eV}$ (SNR threshold=8.0), respectively. For ET in the observation period of 1 year, the bound of graviton mass is $m_g \leq 3.2 \times 10^{-20}\, \mathrm{eV}$ (SNR threshold=8.0), 
which yields almost the same constraints within the $10^{-20}$ orders of magnitude.

In previous research, it is interesting to note that \cite{nishizawa2014} has placed a mass bound of the graviton within the $10^{-20}$ order of magnitude in the absence of positive detection. 
This bound was derived from the study of observations of photons and neutrinos from compact binary mergers and supernovae.
It is important to consider that \cite{nishizawa2014} uniformly chooses the frequency of GWs to be around 
$\sim 100\,$Hz for their calculations, while the frequency of GWs we calculated is almost located at $\sim 700\,$Hz. This difference in frequency could indeed lead to variations in the constraints on the graviton mass.
In addition, the graviton mass has been constrained by several observations of binary pulsars, resulting in a constraint of $m_g<7.6\times 10^{-20}\,\mathrm{eV}$~\citep{graviton2-PhysRevD.65.044022}. Furthermore, \cite{graviton2-PhysRevD.65.044022} has predicted that space-based GW detectors could place better constraint on the graviton mass by observing multiple inspiralling black hole binaries,  potentially leading to a constraint of up to $m_g\simeq 4\times 10^{-26}\,\mathrm{eV}$~\citep{graviton-PhysRevD.84.101501}.
Overall, it is interesting to find that the bound on $m_g$ is comparable with that from binary pulsars,
indicating the consistency and relevance of our results in the context of existing constraints.

\section{Conclusion and Discussion}
\label{sec:conclusion}
In this paper, a simulation study was conducted to constrain the GW propagation speed. The study focused on analyzing the difference in arrival time between GW and sGRB signals from compact binary mergers and considering the distance to the corresponding astrophysical event source.
The simulation involved the joint observation of GW and sGRB signals from a binary neutron star (BNS) merger event using the \texttt{GWToolbox}. The study considered both second and third-generation ground-based GW detectors, specifically the aLIGO in design sensitivity and the ET, for joint detection with the \textit{Fermi}/GBM in different observation settings.

We have found that the joint detection of GWs and sGRB from the BNS merger has provided valuable insights into the constraining potential of aLIGO (in design sensitivity) and ET for the propagation speed of gravitational waves in future multi-messenger observations. 
Applying the MC algorithm, the upper bound of constraint $\delta c/c$ reaches the order of $10^{-17}$ for design aLIGO and $10^{-18}$ for ET,
as indicated in Equation~\eqref{eq: result of bound} and Equation~\eqref{eq: result of bound2}.
Furthermore, the study obtained the coefficient $-4\times 10^{-16}< \bar{s}_{00}^{(4)}<8\times 10^{-18}$ for local Lorentz violation in the SME framework, along with a constraint on the graviton mass of $m_g \leq 7.1(3.2) \times 10^{-20}\,$eV for design aLIGO (ET). 
Both of these values are constrained based on the GW speed constraints from ET.

The results show a significant improvement in the constraint of the GW speed, ranging from one order of magnitude for design aLIGO (over a 20-year observation period) to three orders of magnitude for ET (over a 1-year observation period), compared to the constraint of GW170817 in \cite{gw-grb170817}.
This highlights the advancements in the ability to constrain the GW speed through multi-messenger observations and the potential of advanced GW detectors, especially with the promise of future detectors such as ET.

The bound on the GW propagation speed in this study is relatively conservative, as we simplified our analysis by using the same assumptions as in the calculation of \cite{gw-grb170817} for the upper and lower bound of $\delta c/c$. This involved assuming that the difference in intrinsic emission time of the two signals is 0 seconds for the upper bound and 10 seconds for the lower bound. 
By adopting this approach, we obtained the accuracy of this constraint that can be achieved in future observations. 
It's important to note that our simulations only considered the joint detection of GWs with on-axis sGRB from BNS merger events. However, as more joint detection events of GWs with other signals occur in the future, tighter bounds on the GW speed will be established, underscoring the importance of Multi-Messenger Astronomy in advancing our understanding of fundamental physics.


\normalem
\begin{acknowledgements}
This research is supported by the National Natural Science Foundation of China under grant 12065017. S.-X.Y. acknowledges support from the Chinese Academy of Sciences (grant Nos. E329A3M1 and E3545KU2)

\end{acknowledgements}
  
\bibliographystyle{raa}
\bibliography{main}

\end{document}